# Terawatt-level few-cycle mid-IR pulses through nonlinear self-compression in bulk

V. Shumakova[1], P. Malevich[1], S. Ališauskas[1], A. Voronin[2,3] A.M. Zheltikov[2,3,4],
D. Faccio[5], D. Kartashov[6], A. Baltuška[1,7], A. Pugžlys[1,7]

[1]*Photonics Institute Vienna University of Technology, Gusshausstrasse 27-387, A-1040 Vienna, Austria*
[2]*Physics Department, International Laser Center, M.V. Lomonosov Moscow State University, 119992 Moscow, Russia*
[3]*Russian Quantum Center, ul. Novaya 100, Skolkovo, Moscow Region, 143025 Russia*
[4]*Department of Physics and Astronomy, Texas A&M University, College Station TX, 77843-4242, USA*
[5]*Institute of Photonics and Quantum Sciences, David Brewster Building, DB1.12 Heriot-Watt University Edinburgh, EH14 4AS, UK*
[6]*Institute for Optics and Quantum Electronics, Friedrich-Schiller University Jena, Max-Wien Platz 1, 07743 Jena, Germany*
[7]*Center for Physical Sciences & Technology, Savanoriu Ave. 231 LT-02300 Vilnius, Lithuania.*
pugzlys@tuwien.ac.at

**Abstract: The physics of strong-field applications requires driver laser pulses that are both energetic and extremely short. Whereas optical amplifiers, laser and parametric, boost the energy, their gain bandwidth restricts the attainable pulse duration, requiring additional nonlinear spectral broadening to enable few or even single cycle compression and a corresponding peak power increase. In the mid-IR wavelength range that is critically important for scaling the ponderomotive energy in strong-field interactions, we demonstrate a remarkably simple energy-efficient and scalable soliton-like pulse compression in a mm-long YAG crystal with no additional dispersion management. Sub-three-cycle pulses with >0.65 TW peak power are compressed and extracted before the onset of modulation instability and multiple filamentation as a result of a favorable interplay between strong anomalous dispersion and optical nonlinearity around the wavelength of 3.9 µm. As a striking manifestation of the increased peak power, we show the evidence of mid-infrared pulse filamentation in atmospheric air.**

Ultrashort high energy laser pulses in the 3—8 µm mid-IR spectral range are of particular interest because of their role in high harmonic generation[1-3], filamentation[4-6], unusual regimes for nonlinear optical phenomena[7] and particle acceleration[8]. These applications benefit from the extended oscillation period of the driving electromagnetic field and require, besides the long carrier wavelength, an ultrashort pulse duration of just a few optical cycles and a high pulse energy reaching into the multi-millijoule range. Attaining all three of these conditions simultaneously is one of the primary challenges of ultrafast laser technology. The limited gain bandwidth of optical amplifiers triggered the invention of various types of external pulse compression approaches based on nonlinear-optical spectral broadening and either subsequent or simultaneous dispersion compensation, referred to, respectively, as post-compression and nonlinear self-compression techniques. A powerful post-compression method for the generation of few-cycle pulses in the near IR at the millijoule energy level is based on spectral broadening in a gas-filled hollow waveguide[9-11] occurring in the regime of near-zero or slightly positive dispersion. In the negative dispersion regime, it is possible to realize a soliton-like nonlinear self-compression, whereby the negative group delay dispersion (GDD) of an anomalously dispersive nonlinear medium counteracts the positive GDD arising from self-phase modulation (SPM). The solitonic type pulse compression scenario is widely realized in waveguides and to an extent can also be applied to femtosecond filaments in gas, where a small central fraction of the beam can exhibit self-compression because of the plasma anomalous dispersion[12-15], and to anomalously-dispersive bulk solids[16,17]. For solid media, the energy of the self-compressed pulses in the femtosecond range is limited to low microjoules because the critical power of self-focusing, $P_{cr}$, is at the MW level. For millijoule pulses, relevant to most high-field laser pulse applications, the peak power $P$ exceeds $P_{cr}$ by several orders of magnitude. Propagation of a beam carrying pulses with high $P/P_{cr}$ ratios over an extended distance in a nonlinear medium results in a rapid beam disintegration into multiple filaments and an overall loss of coherence that makes such pulse sources unusable for applications.

Here we report the experimental realization of self-compression in anomalously dispersive transparent solids of femtosecond mid-IR pulses with energies reaching above 20 mJ, which represents a robust and reliable way to produce few-cycle optical pulses with peak powers exceeding 650 GW. The solitonic scenario is verified by extensive 3D numerical modeling which reveals that under our operating conditions, the length of temporal compression is much shorter than the length of spatial modulation instability and beam breakup into multiple filaments. It is this separation of length scales, attainable in the regime of strong anomalous dispersion for our mid-IR pulses, which allows us to use a solid medium for temporal compression at remarkably high energies. Whereas the merits of bulk anomalous dispersion for soliton-like self-compression in the mid-IR have been already duly recognized and exploited in the microjoule pulse range[14,15], this work represents a radical breakthrough by demonstrating that the fundamental limitation imposed by the critical power of self-focusing can be surpassed by multiple orders of magnitudes in the mid-IR spectral range.

Second harmonic generation frequency resolved optical gating (SHG FROG) characterization of 21-mJ 3.9-μm pulses before and after self-compression are shown in Fig.1 with further details described in figures 1 and 2 of the Supplementary Information. The self-compression takes place in a 2-mm thick yttrium aluminum garnet ($Y_3Al_5O_{12}$, YAG) plate (*P*) oriented at the Brewster angle and placed at a distance of 50 cm from the focal plane of a lens with the focal distance of 75 cm. As can be seen from Fig.1b spectrally broadened due to SPM in YAG 94-fs pulses are self-compressed by a factor of 3 to 30 fs (Fig.1c), which at 3.9 μm wavelength corresponds to less than 3 optical cycles. We underscore that the SHG FROG measurement was performed on the whole beam, i.e. without selecting a certain fraction along the transversal coordinate. As seen from the measured dependence of the pulse duration on the input intensity and on the YAG thickness (Fig.1d), self-compression can be achieved in rather broad parameter range. The energy of self-compressed pulses exceeds 19.7 mJ, indicating a >93% combined transmission efficiency.

The origin of the 7% losses was clarified by measuring the dependencies of the transmission on the incident intensity at different material thicknesses (see Supplementary Information). As expected, at low incident intensity <0.1 TW/cm², which was achieved by detuning the compressor, the transmission is 100%. In the case when the compressor is optimized and the input intensity is in the range 0.5 TW/cm$^2$<I<1.5 TW/cm$^2$ the transmission stays at the level of 97% and is independent on the material thickness in the range of 1-3 mm. This reveals that the 3% losses are due to plasma formation in air in the focus of the 75-cm lens. With further increase of the intensity, the transmission decreases with the steeper decrease at larger material thickness, indicating ionization and/or induced absorption losses. This implies only <4% loss in the YAG crystal and reveals that the energy of self-compressed pulses exceeds 20 mJ.

We have also studied the uniformity of the compression across the beam profile. It follows from both the experiments and calculations that the central part of the beam self-compresses more as compared to the overall beam but with an overall variation of the pulse duration that is less than ~30% and across radial coordinate remains very smooth and uniform (as opposed e.g. to the strong pulse splitting observed in filamentation).

The full 3D numerical model (see Methods section) was first used to reproduce the pulse transformation observed in the experiments. Fig.1b,c show the simulation results (contours shaded in yellow) superposed on the experimental results (blue lines). The good quantitative agreement indicates the accuracy of the numerical simulations. The key tendencies of these transformations appear to be very similar to the generic soliton pulse self-compression scenario. However, we underline that the overall picture of pulse compression in a focused high-power beam as implemented in our experiment is much more intricate and cannot be reduced to a one-dimensional textbook soliton dynamics[18]. Most importantly, in a full 3D propagation the spatial dynamics gains critical importance. In typically studied scenarios, spatiotemporal filamentation dominates the pulse dynamics with simultaneous temporal compression and spatial collapse followed either by temporal pulse splitting or by multiple refocusing cycles. Furthermore, at $P$ merely several times over the threshold power of self-focusing, $P_{cr}$, the spatial modulation instability and consequent spatial breakup into multiple filaments becomes imminent with propagation[19]. It is therefore remarkable that under the conditions of our mid-IR pulse experiments, corresponding to the peak power that is four orders of magnitude above $P_{cr}$ for YAG, all these types of detrimental dynamics are reliably suppressed, as is also confirmed by our numerical simulations.

For a quantitative description of the separation of different nonlinear length-scales, we compute the transverse field intensity profiles for different propagation lengths $z$ (Fig.2). For the chosen set of parameters, the shortest pulse width of self-compressing soliton transients is achieved at $z$≈1.45 mm. At this point of maximum pulse compression, the intensity, as can be seen from Fig.2d, reaches its local maximum. Analysis of the beam profile and angular spectrum of the field at this point shows that, although the modulation instabilities start to build up, the beam does not display any noticeable degradation of its intensity profile (Figs.2a-2c). Beyond $z$>3 mm, however, the intensity rapidly raises and is accompanied by the buildup of hot spots across the beam as a result of multiple filamentation, leading to a dramatic degradation of the beam profile and angular spectra (Figs.2a-2c). Consequently, the zone of maximum pulse compression (Fig.2d) is distinctly separated (in propagation distance) from the region where the beam starts to break up into multiple filaments, thus allowing pulse self-compression to be implemented without the detrimental beam break up that should otherwise be expected at such high powers.

As a proof of concept in the strong field regime, directly sensitive to the scaling of the peak power boosted by the nonlinear pulse self-compression, we examined filament formation in air behind a 2-mm thick Brewster-oriented YAG plate placed at different distances from the f=75 cm lens (Fig.3). The observed filamentation in ambient air is a convincing evidence for a dramatically increased peak power of the self-compressed pulse as well as a proof of high pulse and beam fidelity. Furthermore, the position of the center of mass moves towards the focusing lens with increasing self-compression (Fig.3d), which is in agreement with the generally expected dynamics of self-focusing[20,21]. Detailed studies on filamentation of self-compressed mid-IR pulses will be presented in separate publication. Here we would like to stress that the demonstrated simple, virtually loss-free self-compression mechanism of high-energy mid-IR pulses can be straightforwardly employed to radically enhance the performance of longwave drivers for strong-field applications.

**Methods**

**Self-compression experiments.** Sub-100 fs, 21-mJ pulses centered at 3.9 µm at a repetition rate of 20 Hz were generated by a hybrid OPA/OPCPA system [22] based on Type II potassium titanyl arsenate (KTA) nonlinear optical crystals. Experimental setup designed for the investigation of self-compression of 3.9-µm pulses is presented in supplementary materials.

**Filamentation.** Filament photos (Fig.6) were taken with stationary fixed digital photo camera (Canon 350D) with the integration time of 10 s as it is shown in Fig.1a.. For each photo the plate was moved away from the focusing lens with the step of 1 cm. Taken photos were processed by converting into gray-scale color map, performing inverse gamma correction (in order to linearize intensity scale), cropping the area in the vicinity of the filament and integrating along the axis perpendicular to the direction of pulse propagation, which provides longitudinal intensity distribution in a filament (Fig.6c). From the obtained data we have determined the dynamics of the length of the filament (at the level of $1/e^2$) as well as of the position of the center of mass. The position of the center of mass (CoM) was found by the procedure $\frac{\sum_{i=1}^{N} I_i x_i}{\sum_{i=1}^{N} I_i}$, where $I_i$ is the amplitude of the luminescence at the position $x_i$ with $i$ being the number of the pixel.

**Modeling.** To understand the spatiotemporal dynamics of ultrashort pulses behind the self-compression of multi-millijoule femtosecond mid-IR pulses in transparent dielectrics, we performed numerical modeling using the three-dimensional time-dependent generalized nonlinear Schrödinger equation [23,24] for the amplitude of the field, including all the key physical phenomena, such as dispersion and absorption of the medium, beam diffraction, Kerr nonlinearities, pulse self-steepening, spatial self-action phenomena, ionization-induced optical nonlinearities, as well as plasma loss and dispersion. The field evolution equation is solved jointly with the rate equation for the electron density, which includes impact ionization and photoionization with the photoionization rate calculated using the Popov-Perelomov-Terentyev version [25] of the Keldysh formalism [26]. Simulations are performed for typical parameters of YAG crystal – a band gap of 6.4 eV, the Kerr-effect nonlinear refractive index $n_2 = 4\times10^{-16}$ cm$^2$/W, and the higher order Kerr effect (HOKE) coefficient $n_4 = -3\times10^{-29}$ cm$^4$/W$^2$. Dispersion of YAG crystal was included in the model through a Sellmeier relation [27]. Spatial modulation instabilities leading to the formation of multiple filaments are seeded in our model by superimposing a Gaussian-noise modulation on the input beam profile [28]. Simulations were performed using an MPI parallel programming interface on the Chebyshev and Lomonosov supercomputer clusters of Moscow State University. We underline that the model includes all expected nonlinear effects including ionization and yet is able to precisely reproduce our experimental results only if we include a HOKE term. There has been significant discussion in the literature regarding the actual role and relevance of such terms – here the HOKE simply plays the role of an additional saturation term that appeared to be necessary in order to achieve full quantitative agreement with experiments and does not by itself represent a demonstration that HOKE is the mechanism through which this attained.

Figures

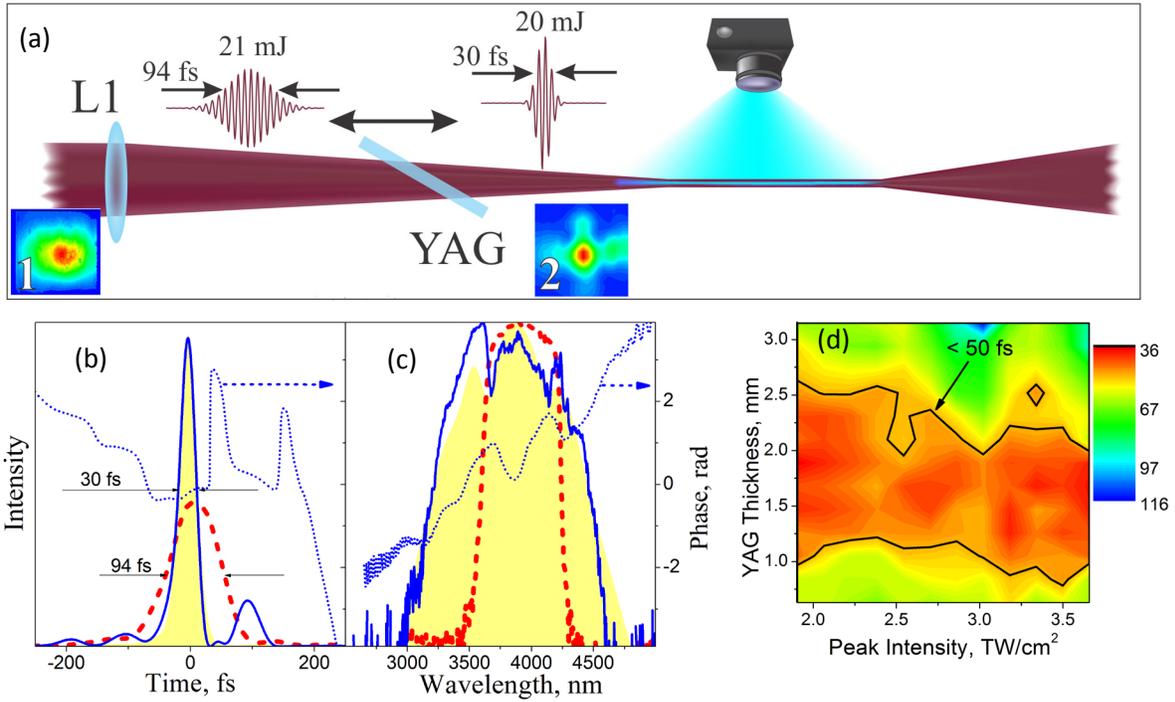

Fig.1. (a): Arrangement of the self-compression setup and beam profiles on the lens (1) and on the output surface of YAG plate (2); (b, c): retrieved from SHG FROG temporal (b) and spectral (c) pulse profiles of the output of 3.9-μm OPCPA system (dashed red line) and self-compressed in YAG pulses (blue solid line). The yellow area represents calculated temporal profile and spectrum of the self-compressed pulse (normalized intensity). Dotted blue lines show retrieved temporal and spectral phases; (d): 3D-map representing the dependence of the output pulse duration on the thickness of the material and on the input peak intensity;

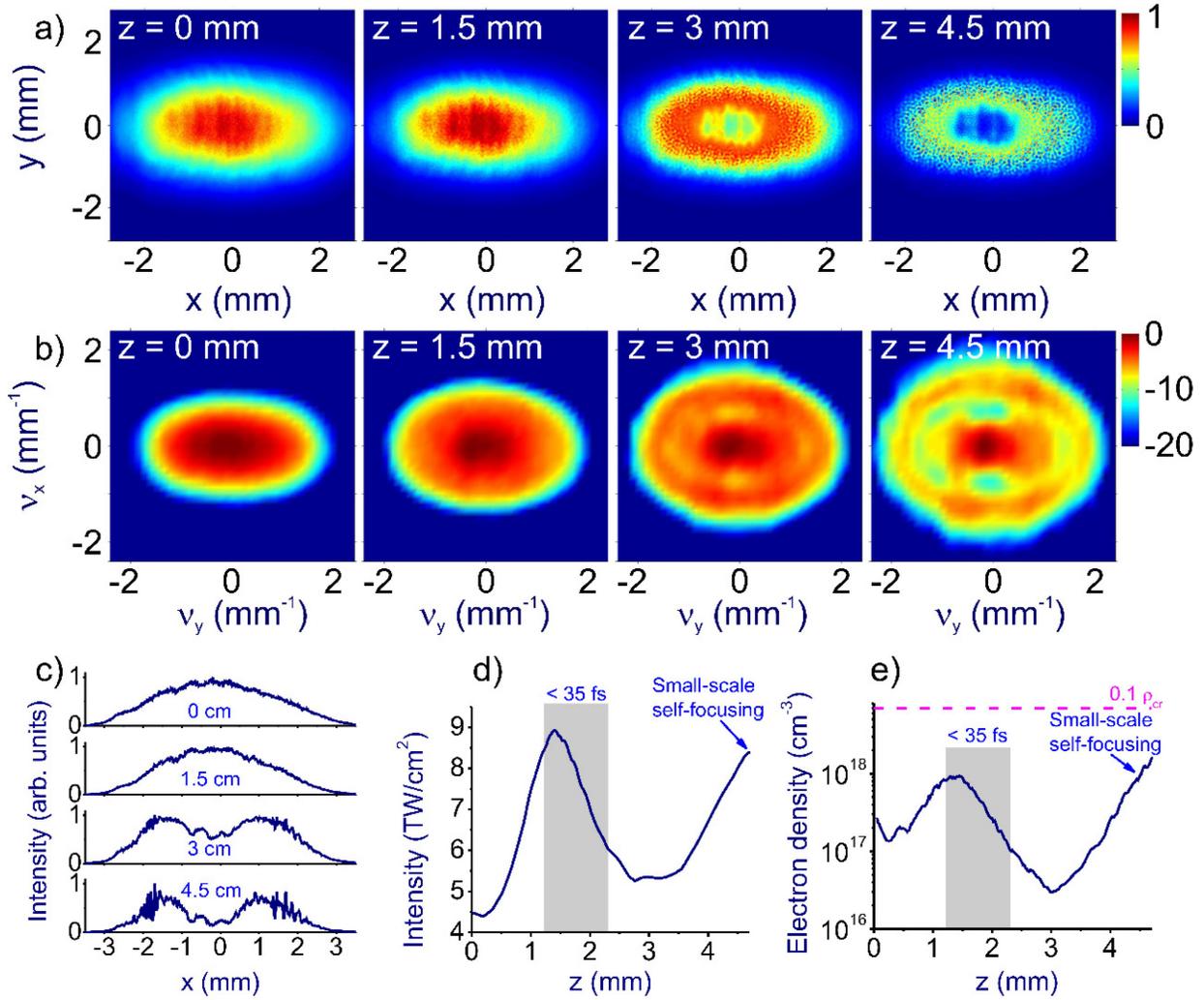

Fig. 2. Three-dimensional simulations for the dynamics of the mid-infrared beam in YAG: (a) beam dynamics, (b) evolution of the angular spectrum, (c) 1D cut of beam dynamics, (d) the field intensity $I_m$, found as the maximum intensity over the beam, and (e) the maximum electron density across the beam calculated as functions of the propagation coordinate $z$.

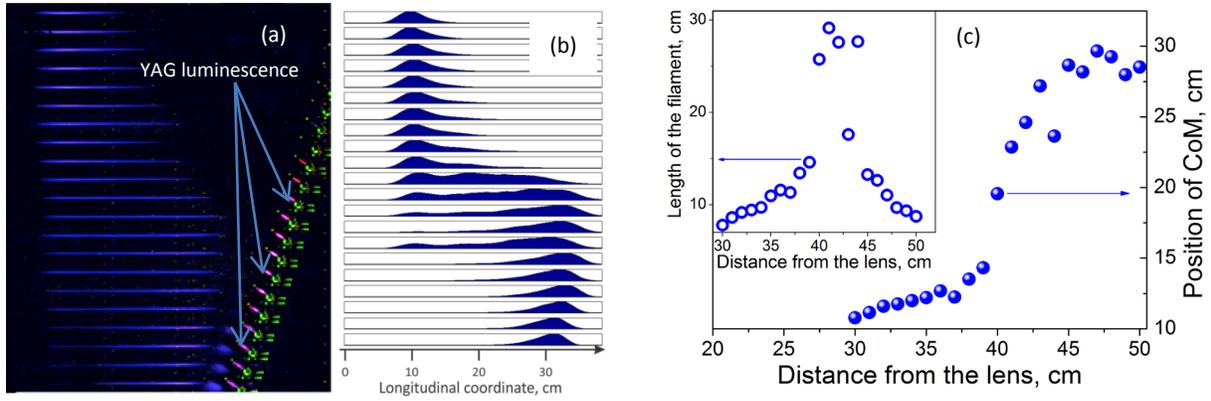

Fig.3 (a): Photos of the visible part of the filament in air at different distance (with a step of 1 cm) between the 75-cm lens and bulk Brewster-oriented 2-mm thick YAG plate. (b): Longitudinal intensity distribution in a filament as extracted after digital processing of the photos. (c): Dependences of the length of the filament (at $1/e^2$ level) and of the position of the center of mas (CoM) on distance between the 75-cm lens and bulk Brewster-oriented 2-mm thick YAG plate.

## Supplementary Figures

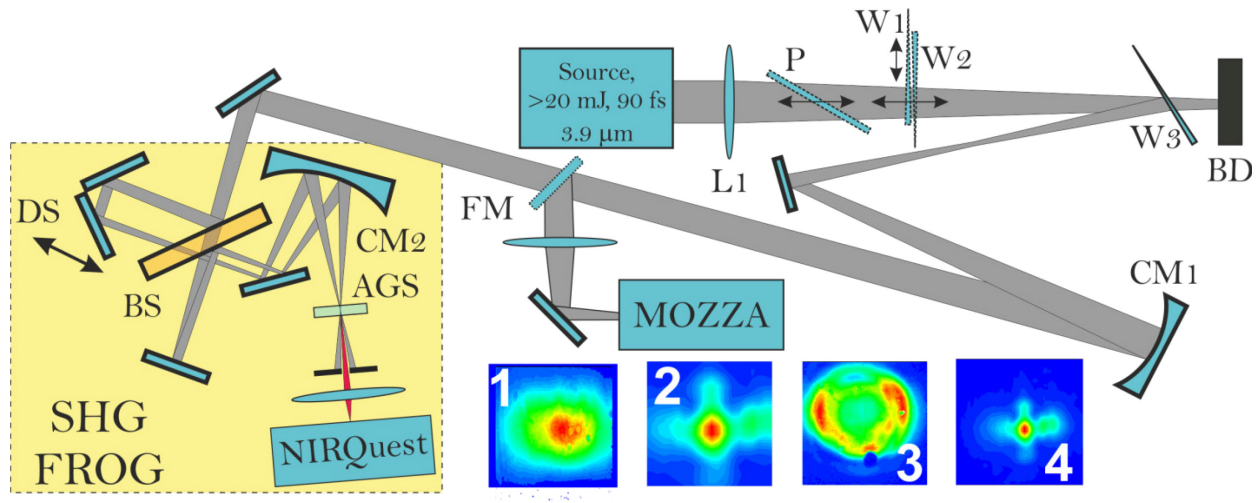

Supplementary Figure 1. Setup for the generation and characterization of self-compressed mid-IR pulses: L1 - $CaF_2$ lens with the focal length of 750 mm; P – 2-mm thick, Brewster angle oriented YAG plate; W1 and W2 - YAG wedges with the apex angle of 6 deg; W3 – $CaF_2$ wedge; BD – beam dump; FM – flip mirror; CM1 and CM2 - spherical mirrors with the radii of curvature 100 cm and 20 cm respectively; MOZZA – acousto-optic spectrometer operating in the spectral range 1-5 µm (FASTLITE); DS – motorized delay stage; BS – broadband pellicle beam-splitter; AGS – 0.2 mm-thick silver thiogallate nonlinear optical crystal; NIR-Quest – near-infrared spectrometer operating in the spectral range 0.9-2.5 µm (OceanOptics); The beam profiles presented at the bottom are taken with a CCD camera (Pyrocam): 1 – at the position of lens L1, 2 – by 2f-2f imaging of the output surface of W2, 3 – at the position of CM1, 4 – at the position of AGS crystal.

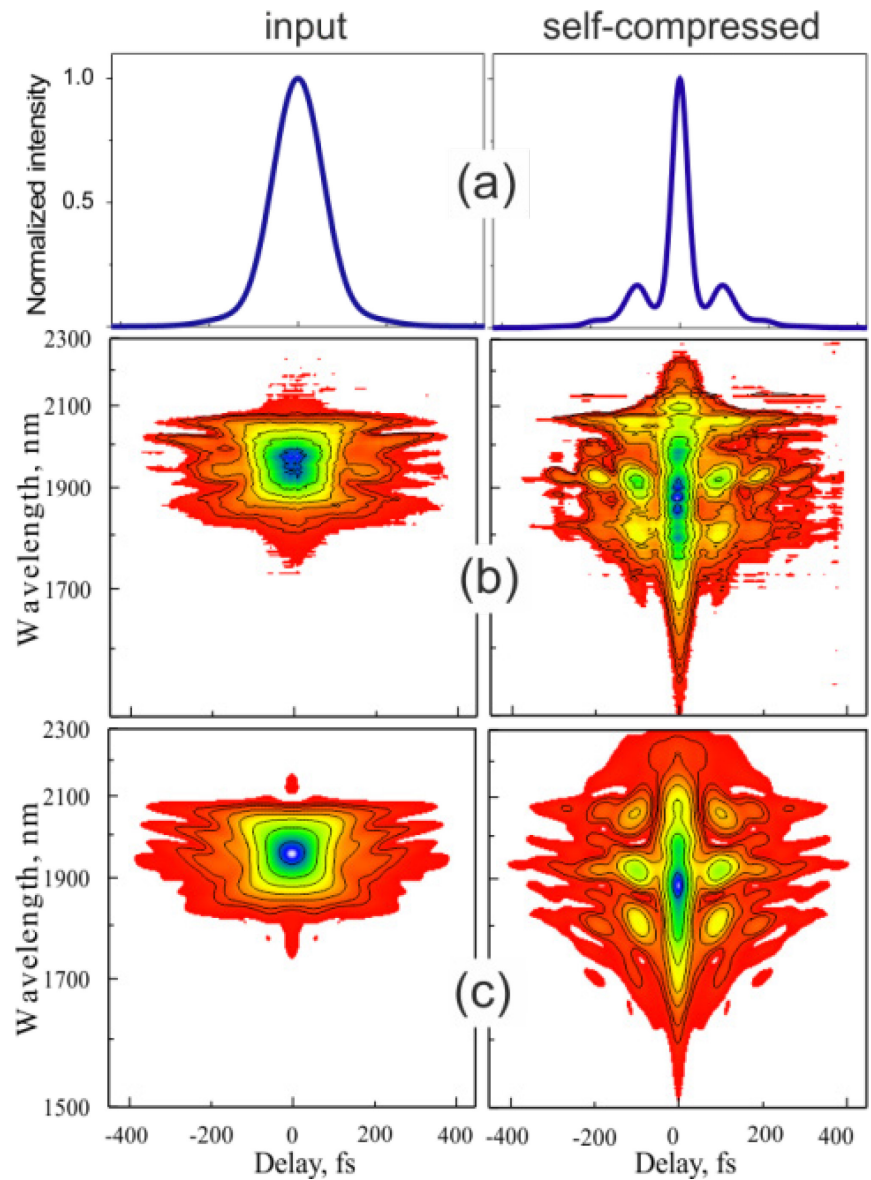

Supplementary Figure 2. Autocorrelation functions (a) and measured (b) and retrieved (c) SHG FROG traces 3.9-μm pulses before (left column) and after (right column) self-compression in YAG.

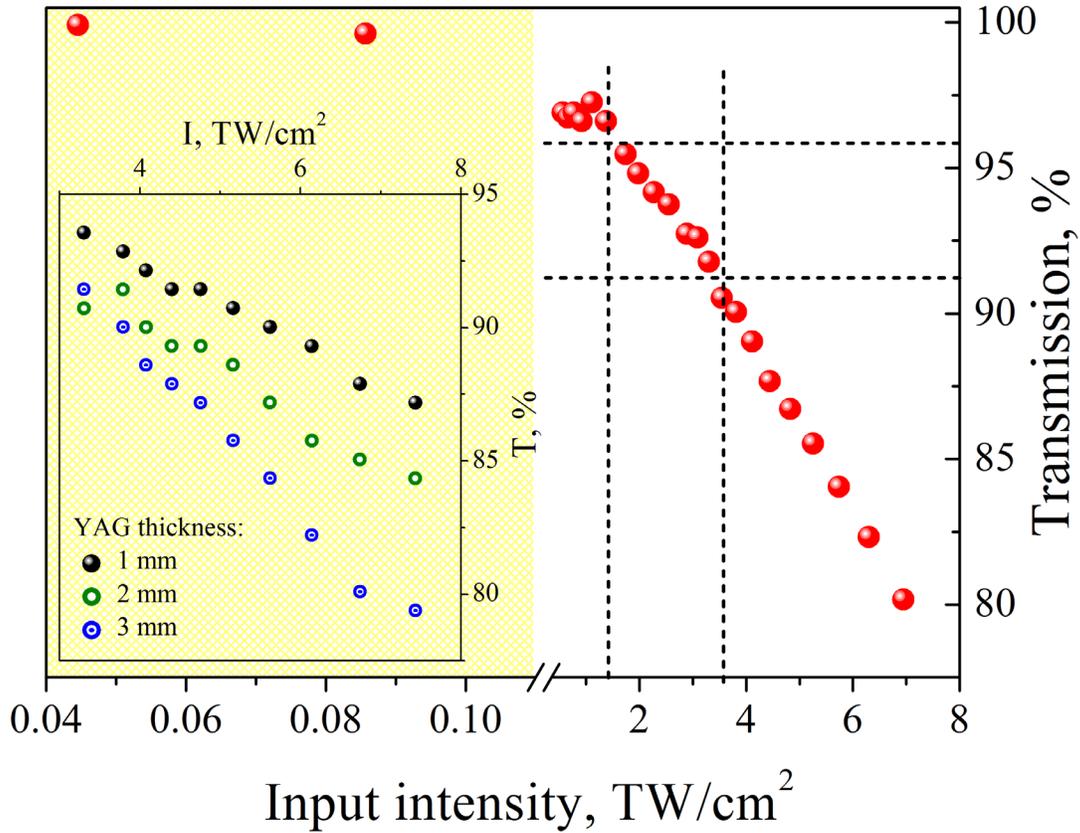

Supplementary Figure 3. Red solid dots: dependence of the optical transmission on the input intensity measured for a 2-mm thick YAG plate at normal incidence; dashed lines bounder the region of efficient self-compression in 2 mm of YAG; the area shaded in yellow displays transmission measurements where the input intensity was controlled by detuning the pulse compressor of the OPCPA from the highest compression ratio; the inset shows the intensity dependent transmission of a pair of YAG wedges for three different thicknesses of material indicated in the figure.

**Supplementary Methods**
**Pulse generation and characterization experimental setup**

Sub-100 fs pulses centered at 3.9 µm at a repetition rate of 20 Hz were generated by a hybrid OPA/OPCPA system based on Type II potassium titanyl arsenate (KTA) nonlinear optical crystals. The three-stage OPA and three-stage OPCPA are respectively pumped by 1-mJ, 200-fs Yb:CaF$_2$ and 1-J, 100-ps Nd:YAG laser systems. Both femtosecond and picosecond laser amplifiers are seeded by a common Yb:KGW laser oscillator, which ensures all-optical synchronization of the OPA and OPCPA units. The OPCPA is seeded by stretched in a GRISM stretcher signal pulses centered at 1460 nm, the idler 3.9-µm pulses generated in the second and amplified in the third OPCPA stage are compressed to sub-100 fs pulse duration in a diffraction-grating based compressor. The energy of compressed pulses exceeds 20 mJ.

Experimental setup designed for the investigation of self-compression of 3.9-µm pulses is presented in Supplementary Figure 1. Compressed pulses were focused by a plano-convex CaF$_2$ lens *L1* with the focal length of 75 cm. A pair of YAG wedges (*W1, W2*) with the apex angle of 6º or, alternatively, a 2-mm-thick YAG window oriented at Brewster's angle were inserted into the beam. The intensity in YAG was varied by changing the distance between the wedge-pair (or the YAG window) and the lens *L1*. The thickness of YAG material was controlled in a continuous way by varying the insertion depth of the wedges. Self-compressed pulses were characterized by the technique of the second-harmonic-generation frequency-resolved optical gating (SHG FROG) based on a Type I 0.2 mm-thick silver thiogallate (*AGS*) nonlinear optical crystal and a near-infrared spectrometer (NIRQuest, OceanOptics). Two pulse replicas in the SHG FROG apparatus were produced by a broadband 2-µm-thick pellicle beam splitter (Thorlabs). To equalize the spectra of both replicas, the beams were once transmitted through and once reflected by the beam splitter. Pulses directed to the SHG FROG were attenuated by reflecting approximately 1.5% of the total energy by a CaF$_2$ wedge *W3*. Spatial-temporal distortions of the self-compressed pulses were circumvented by a 4-f imaging of the output surface of the wedge *W2* (YAG window) on the surface of AGS crystal which was performed by a pair of spherical mirrors *CM1* and *CM2* with the radii of curvature of 100 cm and 20 cm. The 4-f imaging resulted in 5× reduced size of the spot on the AGS crystal as compared to the size on the *W2*(YAG window). The spectra of mid-IR pulses were recorded by acousto-optic based scanning spectrometer (MOZZA, FASTLITE).

**Determination of the origin of losses**

The origin of the 7% losses was clarified by measuring the dependencies of the transmission on the incident intensity at different material thicknesses (Supplementary Figure 3). In order to have an adjustable material thickness we used a pair of YAG wedges (*W1* and *W2*), shown in Supplementary Figure 1, the YAG wedges were placed at close to normal incidence with respect to the laser beam to have maximum accessible range of the material thickness. Supplementary

Figure 3 shows the transmission dependence corrected for the Fresnel losses on the YAG surfaces (n=1.757). As expected, at low incident intensity below 0.1 TW/cm$^2$, which was achieved by detuning the compressor, the transmission is 100%. In the case when the compressor is optimized and input intensity is in the range 0.5 TW/cm$^2$<$I$<1.5 TW/cm$^2$ the transmission stays at the level of 97% and is independent on the material thickness in the range of 1-3 mm. This reveals that the 3% losses are due to plasma formation in air in the focus of the 75-cm lens. With further increase of the intensity, the transmission decreases with the steeper decrease at larger material thickness (inset in Supplementary Figure 3), indicating ionization and/or induced absorption losses. This implies only <4% loss in the YAG crystal and reveals that the energy of self-compressed pulses exceeds 20 mJ.